\documentclass[aps,prb,reprint,preprintnumbers,amsmath,amssymb,showpacs,superscriptaddress,citeautoscript]{revtex4-2}
\usepackage[T1]{fontenc}
\usepackage{amssymb}

\usepackage{graphicx}

\usepackage{amsmath}
\usepackage{bm}
\usepackage{color, soul}
\usepackage[normalem]{ulem}
\usepackage{siunitx}
\usepackage{upgreek}
\usepackage[dvipsnames]{xcolor}
\usepackage{comment}
\usepackage{natbib}
\usepackage{braket}
\usepackage{ulem}

\usepackage[breaklinks]{hyperref}
\usepackage[all]{hypcap}
\usepackage[all]{hypcap}
\hypersetup{
    plainpages=false,
    unicode=false,          
    pdfmenubar=true,        
    pdffitwindow=false,     
    pdfstartview={FitH},    
    pdftitle={Geometronics},    
    pdfauthor={Maciej Śmiertka},     
    pdfproducer={PWr}, 
    pdfkeywords={High} {Magnetic} {Fields}, 
    pdfnewwindow=true,      
    linktoc=section,
    colorlinks=true,       
    linkcolor=blue,          
    citecolor=red,        
    filecolor=magenta,      
    urlcolor=blue           
}

\DeclareUnicodeCharacter{2212}{-}

\usepackage{xr}

\begin{document}


\title{Geometry-Controlled Magnetic and Electronic Landscapes in Anisotropic van der Waals Materials}

\author{Maciej Śmiertka}
\thanks{These authors contributed equally to this work.}
\affiliation{Department of Experimental Physics, Faculty of Fundamental Problems of Technology, Wroclaw University of Science and Technology, 50-370 Wroclaw, Poland}

\author{Ewelina Cybula}
\thanks{These authors contributed equally to this work.}
\affiliation{Department of Experimental Physics, Faculty of Fundamental Problems of Technology, Wroclaw University of Science and Technology, 50-370 Wroclaw, Poland}

\author{Oliwia Janikowska}
\affiliation{Department of Experimental Physics, Faculty of Fundamental Problems of Technology, Wroclaw University of Science and Technology, 50-370 Wroclaw, Poland}

\author{Bartosz Hołyński}
\affiliation{Department of Experimental Physics, Faculty of Fundamental Problems of Technology, Wroclaw University of Science and Technology, 50-370 Wroclaw, Poland}

\author{Gayatri}
\affiliation{Institute of Experimental Physics, Faculty of Physics, University of Warsaw, Pasteura 5, 02-093, Warsaw, Poland}

\author{Grzegorz Krasucki}
\affiliation{Institute of Experimental Physics, Faculty of Physics, University of Warsaw, Pasteura 5, 02-093, Warsaw, Poland}

\author{Mariusz Hasiak}
\affiliation{Department of Mechanics, Materials and Biomedical Engineering, Wrocław University of Science and Technology, Wroclaw 50-370, Poland}

\author{Kseniia Mosina}
\affiliation{Department of Inorganic Chemistry, University of Chemistry and Technology Prague, Technicka 5, Prague 6, 16628 Czech Republic}

\author{Zdenek Sofer}
\affiliation{Department of Inorganic Chemistry, University of Chemistry and Technology Prague, Technicka 5, Prague 6, 16628 Czech Republic}

\author{Adam Babiński}
\affiliation{Institute of Experimental Physics, Faculty of Physics, University of Warsaw, Pasteura 5, 02-093, Warsaw, Poland}

\author{Maciej R Molas}
\affiliation{Institute of Experimental Physics, Faculty of Physics, University of Warsaw, Pasteura 5, 02-093, Warsaw, Poland}

\author{Paulina Plochocka}\email{paulina.plochocka@lncmi.cnrs.fr}
\affiliation{Department of Experimental Physics, Faculty of Fundamental Problems of Technology, Wroclaw University of Science and Technology, 50-370 Wroclaw, Poland}
\affiliation{Laboratoire National des Champs Magn\'etiques Intenses, EMFL, CNRS UPR 3228, Universit{\'e} Grenoble Alpes, Universit{\'e} Toulouse, Universit{\'e} Toulouse 3, INSA-T, Grenoble and Toulouse, France}

\author{Micha{\l} Baranowski}\email{michal.baranowski@pwr.edu.pl}
\affiliation{Department of Experimental Physics, Faculty of Fundamental Problems of Technology, Wroclaw University of Science and Technology, 50-370 Wroclaw, Poland}

\date{\today}

\begin{abstract}
\noindent Electronic structure in van der Waals materials is commonly engineered through composition, strain, electrostatic gating and heterostructure assembly. Here we introduce geometronics, a concept in which substrate geometry locally reorients an anisotropic crystal, transforming homogeneous external perturbation into programmable magnetic and electronic landscapes. We demonstrate this concept using a bilayer of the antiferromagnetic semiconductor CrSBr transferred onto an inverted pyramidal nanoindentation, where the local crystal orientation with respect to the external magnetic field drives the coexistence of antiferromagnetic and ferromagnetic phases within a single continuous crystal. The resulting magnetic landscape creates a switchable excitonic potential well of up to 10--12 meV, directly visualised by spatially resolved spectroscopy. More generally, geometronics provides a universal route for deterministically programmed electronic and magnetic landscapes without modifying the material itself. It therefore establishes substrate topography as a new design principle that exploits the intrinsic anisotropy of layered van der Waals materials.
\end{abstract}

\maketitle

\begin{figure*}[]
    \centering
    \includegraphics[width=\textwidth]{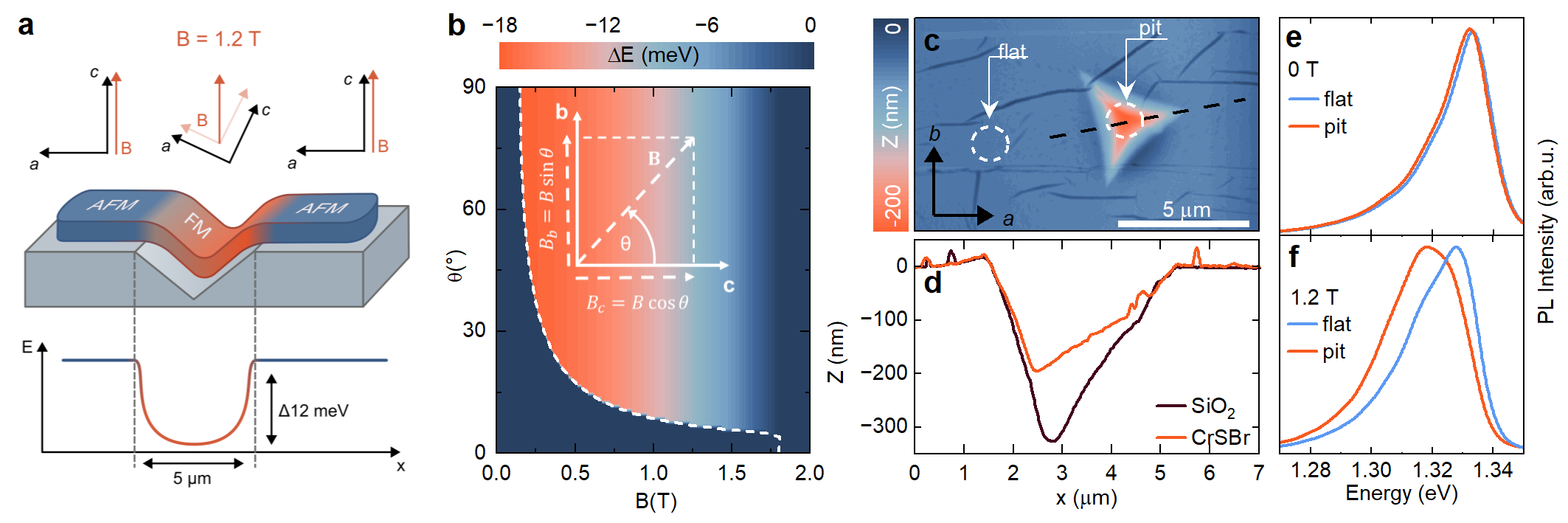}
   \caption{{\bf Concept of geometronics and sample topography with magneto-optical characterization of a CrSBr bilayer transferred onto a nanoindentation pit.} {\bf a}, Schematic illustration of a CrSBr bilayer conforming to the substrate topography under an external magnetic field applied along the crystallographic $c$-axis of the flat region, showing the emergence of a local ferromagnetic phase in the reoriented part of the flake together with the corresponding excitonic energy landscape. {\bf{b}}, Calculated energy difference between the excitonic ground-state transition of the flat and reoriented regions of the CrSBr flake, $\Delta E = E_\mathrm{flat} - E_\mathrm{pit}$, as a function of the applied magnetic field for field rotations within the $bc$ crystallographic plane. The magnetic-field orientation is parameterized by the angle $\theta$, defined in the sketch. {\bf c}, Atomic force microscopy image of the CrSBr bilayer transferred onto a Si/SiO$_2$ substrate containing an inverted pyramidal nanoindentation. The dashed line indicates the direction along which the representative height profile was extracted. {\bf d}, Height profiles measured by atomic force microscopy across the nanoindentation before (black) and after (orange) transfer of the CrSBr bilayer. {\bf e, f}, Low-temperature ($\sim$10 K) photoluminescence spectra acquired from the flat and reoriented regions under magnetic fields of 0 and 1.2 T, respectively, applied along the crystallographic $c$-axis of the flat region. Blue and orange curves correspond to the flat and reoriented regions, respectively.}
    \label{fig:Fig1}
\end{figure*}

\noindent Layered van der Waals materials have opened a new chapter in the engineering of the electronic structure of solids. Their weak interlayer bonding enables atomically thin crystals to be isolated and combined into arbitrary heterostructures \cite{geim2013van, castellanos2022van} with controlled twist angle \cite{sun2024twisted}, or integrated with virtually any substrate providing local strain control \cite{iff2017substrate,jasinski2022strain,kumar2015strain,singh2025quantum, kern2016nanoscale,paralikis2025tunable}, pushing materials design far beyond the limits reached by conventional epitaxy.  Such flexibility allows exploration of new physical phenomena, exemplified by Moiré physics \cite{tran2019evidence,seyler2019signatures,gu2022dipolar,huber2026optical, kennes2021moire, cao2018unconventional, li2021lattice, xu2023observation} or local strain-engineered electronic properties \cite{branny2016deterministic,rosati2021dark,cenker2022reversible,kumar2015strain, linhart2019localized, jasinski2022strain}. Yet layered materials offer another largely unexplored opportunity. Their intrinsic crystal anisotropy (distinguishing in-plane and out-of-plane directions) makes their electronic, optical, and magnetic responses strongly dependent on the direction of the external perturbation, providing a fundamentally different degree of freedom to control electronic structure. 

Despite the remarkable success of current engineering strategies, the electronic landscape is largely determined during the fabrication process. Once a heterostructure is assembled or a strain profile is imposed, the spatial distribution of electronic properties becomes essentially fixed. Dynamic, on-demand tuning is then typically achieved by applying spatially uniform external stimuli, such as electric or magnetic fields, which, however, modify the entire crystal in the same manner.

Here we introduce a fundamentally different approach to local control of electronic structure, in which the intrinsic anisotropy of the crystal is used to transform a spatially uniform external field into a position-dependent perturbation. Owing to their exceptional mechanical compliance, atomically thin crystals readily conform to non-planar substrates while preserving their local crystallinity \cite{branny2016deterministic,paralikis2025tunable,qi2023recent,jasinski2022strain}. As a result, a layer deposited onto a three-dimensional surface acquires a continuously varying crystallographic orientation with respect to the laboratory reference frame. Consequently, even a spatially uniform external field is projected differently onto the local crystal axes at different positions across the layer, producing position-dependent modifications of the electronic structure. 

We demonstrate this concept using the layered antiferromagnetic semiconductor CrSBr \cite{ziebel2024crsbr, klein2023bulk}, whose pronounced magnetic anisotropy \cite{goser1990magnetic, telford2020layered} and strong coupling between magnetic order and electronic structure \cite{wilson2021interlayer,dirnberger2023magneto,smiertka2026distinct} provide an ideal model system. Placing thin CrSBr flakes on substrates with controlled topography, we create spatially varying crystal orientations within a continuous layer, enabling on-demand control of the local electronic structure and magnetic phase using a uniform external magnetic field. Using spatially resolved magneto-optical spectroscopy, we directly visualize these local modifications arising solely from the interplay between geometry and the applied field.\\

\noindent\textbf{Geometronics with CrSBr}

\noindent CrSBr provides an ideal platform to implement the concept of geometronics because its electronic structure is strongly coupled to the magnetic state. This layered semiconductor exhibits A-type antiferromagnetic (AFM) order \cite{goser1990magnetic,ziebel2024crsbr,telford2020layered}, and magnetic-field-induced transitions between the AFM and ferromagnetic (FM) phases are accompanied by pronounced modifications of the electronic structure, and thus shifts of the ground excitonic transition by $\sim12-18$ meV \cite{wilson2021interlayer, dirnberger2023magneto,smiertka2026distinct,komar2024colossal, Antoniazzi2026_CrSBr}. At the same time, the orthorhombic crystal structure gives rise to a highly anisotropic magnetic response \cite{goser1990magnetic,telford2020layered,lee2021magnetic,ziebel2024crsbr}. In the AFM ground state, spins are aligned along the magnetic easy b-axis, where an external magnetic field induces a spin-flip transition at a critical field of approximately 0.15~T \cite{wilson2021interlayer} (at a temperature of around 5\,K). In contrast, fields applied along the a- and c-axes induce a continuous canting of the spins towards the field direction, with characteristic critical fields of approximately 0.8 T and 1.7 T, respectively \cite{wilson2021interlayer}.

\begin{figure*}[]
    \centering
    \includegraphics[width=1\textwidth]{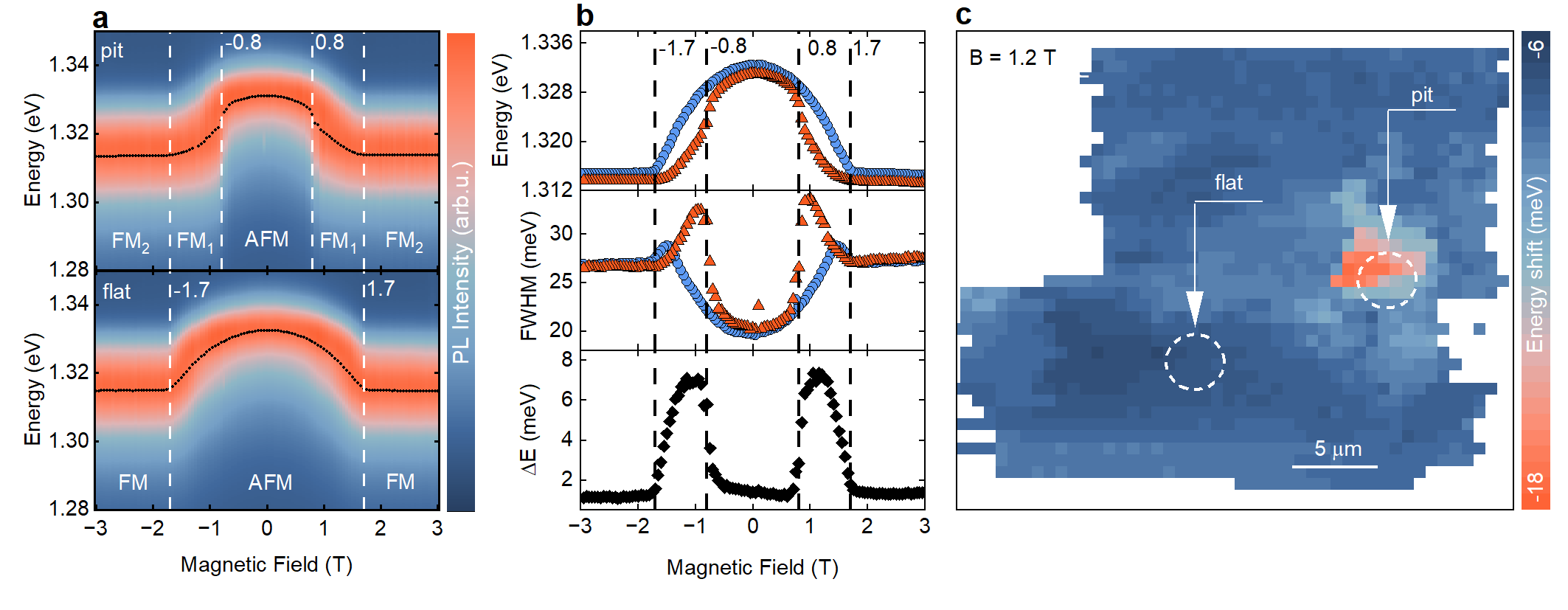}
    \caption{{\bf Magnetic-field control of the excitonic landscape.}  {\bf a}, False-colour maps showing the evolution of the low-temperature ($\sim$10 K) PL spectra with magnetic field for the flat and pit regions of the CrSBr bilayer. The magnetic field is applied along the crystallographic $c$-axis of the flat region. Black symbols mark the PL peak position. {\bf b}, Magnetic-field dependence of the PL peak energy (top), PL linewidth (middle), and energy difference between the reoriented and flat regions, $\Delta E = E_\mathrm{flat} - E_\mathrm{pit}$ (bottom). In the top and middle panels, blue circles and orange triangles correspond to the flat and pit regions, respectively. {\bf c} Spatial map of the photoluminescence (PL) energy shift between 0 and 1.2~T, ($ E_{\mathrm{shift}} = E_\mathrm{0T} - E_\mathrm{1.2T}$) revealing the formation of a localised excitonic potential within the nanoindentation.}
    \label{fig:Fig2}
\end{figure*} 

Such a pronounced anisotropy allows geometry to locally control the magnetic order. When a thin CrSBr flake conforms to a substrate with a designed topography, the local orientation of the crystal axes varies across the sample (Fig.~\ref{fig:Fig1}a), so that a uniform out-of-plane field acquires position-dependent projections onto the local $a$-, $b$- and $c$-axes. Because these axes possess distinct magnetic responses, neighbouring regions evolve through the magnetic phase diagram at different rates and can host distinct configurations coexisting at a given field, ranging from differently canted AFM states to fully ferromagnetic order (see extended discussion in SI and Fig.\,S1 for details). Since each magnetic state has a different electronic structure, this spatial variation translates directly into a position-dependent bandgap and exciton landscape, as illustrated schematically in Fig.~\ref{fig:Fig1}a. 

To provide a quantitative framework for the geometry-controlled electronic landscape, we developed a simple phenomenological model describing the evolution of the magnetic configuration and the corresponding excitonic transition energy in CrSBr under an arbitrarily oriented magnetic field (see Supporting Information and Fig. S2). As an example, Fig.~\ref{fig:Fig1}b presents the calculated energy difference between the excitonic transition measured in the flat region of the flake ($\theta=0^\circ$) and in a locally reoriented region ($\theta\neq0^\circ$), for a magnetic field rotating within the $bc$ plane. The angle $\theta$, defined with respect to the crystallographic $c$ axis, is illustrated schematically in Fig.~\ref{fig:Fig1}b.

The calculated phase diagram demonstrates that even a modest local reorientation of the crystal ($\theta\approx10^\circ$--$20^\circ$) is sufficient to generate an excitonic energy contrast exceeding 10~meV at magnetic fields below 1~T. This regime corresponds to the onset of the spin-flip transition in the reoriented region, where the magnetic-field component along the local $b$ axis exceeds the critical value required for ferromagnetic alignment, while the flat region remains in the canted antiferromagnetic state. As the magnetic field is further increased, the energy contrast gradually decreases because the spins in the flat region continuously cant towards the ferromagnetic configuration, reducing the difference in magnetic order between the two regions. Importantly, this phase diagram demonstrates that the pronounced anisotropy of the magnetic response in CrSBr makes the geometronic effect robust over a broad range of crystal orientations and magnetic fields, allowing substantial electronic contrast to be achieved even for relatively small substrate inclination angles.\\

\begin{figure*}[]
    \centering
    \includegraphics[width=1\textwidth]{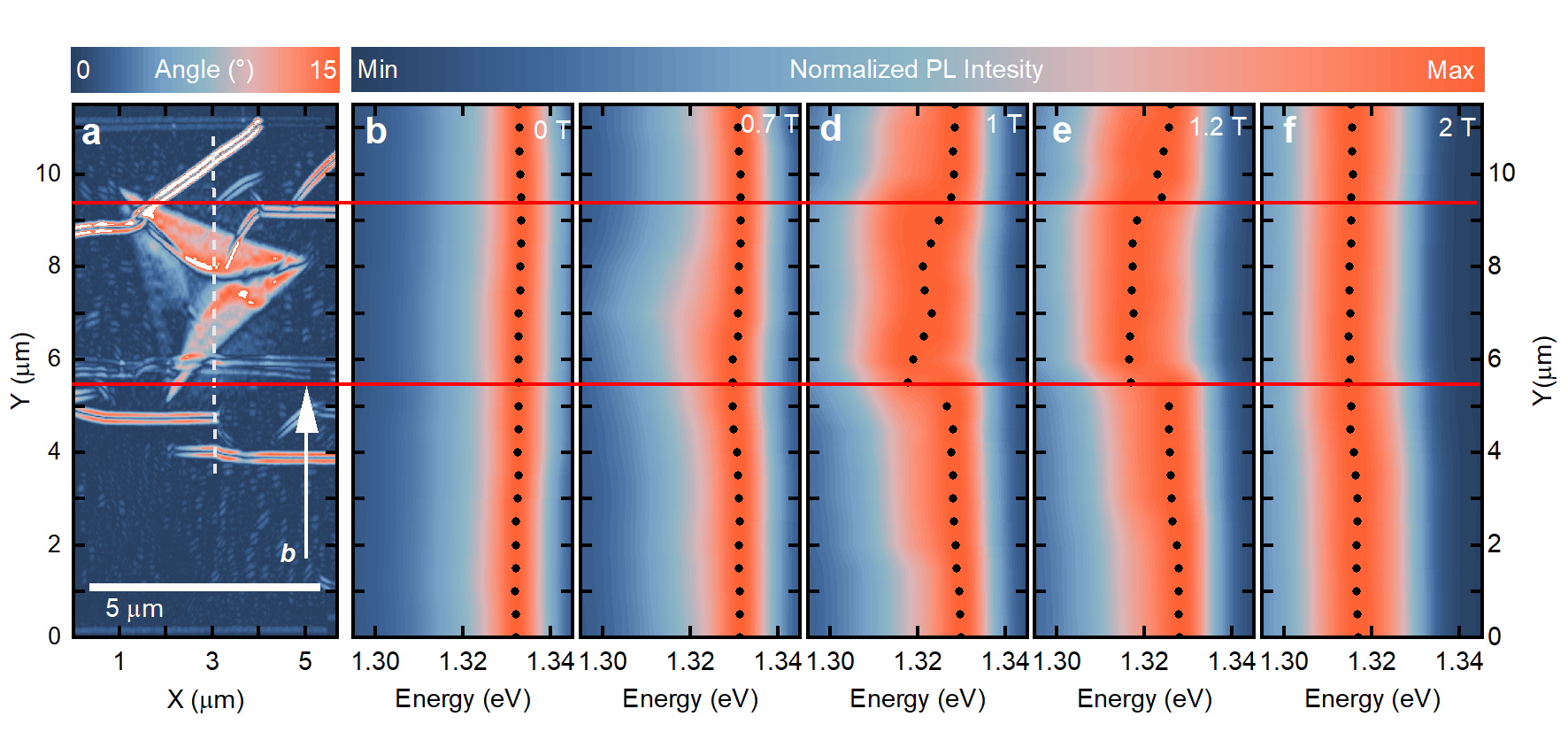} 
    \caption{{\bf Spatially resolved excitonic landscape.} {\bf a}, False-colour map of the absolute local tilt angle of the CrSBr bilayer extracted from the atomic force microscopy topography, measured along the crystallographic $b$-axis with respect to its orientation in the flat region of the substrate. The vertical dashed line indicates the position of the photoluminescence (PL) line scan shown in {\bf b-f}, while the horizontal solid red lines mark the lateral boundaries of the nanoindentation. {\bf b–f}, False-colour maps of the low-temperature ($\sim$10 K) PL spectra acquired along the crystallographic $b$-axis under magnetic fields of 0, 0.7, 1.0, 1.2 and 2.0 T applied along the crystallographic $c$-axis of the flat region. Black symbols indicate the PL peak energy, directly visualizing the spatial dependent evolution of the geometry-programmed excitonic landscape.}
    \label{fig:line_scans}
\end{figure*}

\noindent\textbf{Magnetic-field control of the excitonic landscape}

\noindent Bilayer CrSBr flakes were transferred onto inverted triangular pyramidal pits fabricated in a SiO$_2$ substrate using nanoindentation (see Methods for details). The resulting topography, measured by atomic force microscopy, is shown in Fig.~\ref{fig:Fig1}c. The cross-section extracted along the dashed line (Fig.~\ref{fig:Fig1}d) demonstrates that the bilayer mostly follows the substrate profile rather than bridges over the structure. The indentation has a lateral size of 4.8~$\mu$m and a depth of 320~nm, corresponding to maximum sidewall inclination angles of $10$--$15^\circ$ with respect to the substrate plane, which rotate the local crystal axes by the same amount relative to the flat region. Such moderate inclination angles are particularly advantageous for the magneto-photoluminescence measurements, as they maintain efficient optical access to the reoriented regions in the back-scattering geometry despite the in-plane orientation of the optical transition dipole in CrSBr~\cite{dirnberger2023magneto,wilson2021interlayer}. At the same time, as predicted by the phenomenological phase diagram (Fig.~\ref{fig:Fig1}b), this range of inclination angles is sufficient to generate a pronounced electronic contrast between the flat and reoriented regions.

In the absence of a magnetic field, the PL response inside and outside the pit does not change significantly, despite the different local geometry, as shown in Fig.~\ref{fig:Fig1}e. This observation immediately indicates that the substrate topography itself does not meaningfully modify the electronic structure, excluding the presence of strain \cite{cenker2022reversible}. The absence of significant strain is further corroborated by spatially resolved Raman measurements (Fig. S3). In striking contrast, the application of a magnetic field produces a pronounced difference between the two regions, as shown in Fig.~\ref{fig:Fig1}f. The excitonic transition associated with the inclined region redshifts by approximately 13--15~meV, consistent with the transition to the ferromagnetic state \cite{wilson2021interlayer, smiertka2026distinct, komar2024colossal,dirnberger2023magneto}. In comparison, the same transition in the flat region shifts by only about 6~meV, indicating that the magnetic moments remain in a partially canted antiferromagnetic configuration rather than reaching full ferromagnetic alignment. These observations demonstrate that antiferromagnetic and ferromagnetic phases can coexist within a single monocrystalline CrSBr flake solely as a consequence of its local geometry under a spatially uniform magnetic field, indicating that magnetic domains can be created and programmed on demand without modifying the material itself.



To further quantify this behaviour, we compare magnetic-field-dependent photoluminescence spectra collected from the flat and inclined regions of the flake (Fig.~\ref{fig:Fig2}a, b). Both regions show the characteristic redshift of the excitonic transition with increasing field, but their evolution differs markedly. In the flat region, the transition energy follows a nearly parabolic dependence associated with spin canting towards the crystallographic $c$-axis, saturating above a critical field of approximately 1.7~T. In the inclined region, an initial parabolic redshift is observed only up to approximately 0.8~T, above which the transition energy rapidly approaches its ferromagnetic value.

This behaviour is a direct consequence of the local crystal orientation within the pit. Whereas the flat region experiences the magnetic field exclusively along the crystallographic $c$-axis, the inclined facets expose all three crystallographic axes to finite field projections as shown in Fig.~S4. Consequently, the low-field evolution is governed by the gradual canting of the spins induced by the field components along the $c$ and/or $a$-axes. As the projection of the vertically applied magnetic field onto the local $b$-axis reaches the spin-flip critical field of $\sim$0.15~T \cite{wilson2021interlayer}, neighbouring layers undergo a rapid antiferromagnetic-to-ferromagnetic transition. The associated enhancement of interlayer electronic hybridisation \cite{wilson2021interlayer, smiertka2026distinct, telford2022coupling, dirnberger2023magneto} drives the abrupt redshift of the excitonic transition visible in Fig.~\ref{fig:Fig2}b and c. Notably, the onset of this transition at an applied vertical magnetic field of approximately 0.8~T is in quantitative agreement with the measured local tilt of the crystal along the $b$-axis (Fig.~S4).


The origin of this abrupt magnetic transition is further corroborated by the evolution of the photoluminescence linewidth. Around 0.8~T, the full width at half maximum increases by more than 50\%, indicating the coexistence of distinct magnetic phases with different excitonic transition energies within the excitation spot. The broadening reflects the distribution of local crystal orientations inside the pit, causing different facets to undergo the magnetic transition at slightly different applied magnetic fields. Finally, the energy difference between the flat and inclined regions disappears above 1.7~T, once the entire crystal reaches the ferromagnetic state. The maximum excitonic energy difference is achieved between 0.8--1.2~T, where it is approximately 7~meV (Fig.~\ref{fig:Fig2}b lowest panel).

The more detailed, spatially resolved photoluminescence map measured at 1.2~T presented in Fig.~\ref{fig:Fig2}c  demonstrates that at the most strongly reoriented positions the exciton confinement can be even higher and reach $\sim 12$\,meV (PL redshift of up to $\sim 18$\,meV in indentation areas while outside only by 6\,meV), in reasonable agreement with our effective model. Importantly, the enhanced redshift characteristic of the ferromagnetic phase is confined entirely to the indentation region (Fig.~\ref{fig:Fig2}c). In contrast, the surrounding flat region exhibits a substantially smaller and uniform redshift of approximately 6~meV, characteristic of the canted antiferromagnetic state. Evidently, the observed excitonic energy landscape directly reflects the local geometry of the crystal under a spatially uniform magnetic field.

This field-dependent energy contrast is further resolved in space by photoluminescence line scans acquired across the pit along the crystallographic $b$-axis (Fig.~\ref{fig:line_scans}). For this scan direction, the local tilt primarily generates a magnetic-field projection onto the magnetic easy axis (see panel a of Fig.~\ref{fig:line_scans}), maximising the contrast between the flat and inclined regions at intermediate magnetic fields. Consistent with the magnetic field-dependent PL, presented in Fig. \ref{fig:Fig2}, negligible energy contrast is observed at 0~T and 2~T, whereas a pronounced, spatially localised energy modulation and linewidth broadening develops exclusively within the pit at intermediate magnetic fields (0.7--1.2~T). The largest contrast is observed around 1.2~T, where the excitonic transition exhibits a maximum redshift of approximately 10~meV relative to the surrounding flat region. Notably, the transition energy is not uniform within the pit itself but exhibits distinct local variations arising from the substrate topography and the local conformation of the CrSBr flake. The well-defined spectroscopic contrast between neighbouring regions indicates that the interfaces separating these phases remain sharp. Although our spatial resolution ($\sim1~\mu$m) does not allow their width to be determined directly, these measurements suggest that the interfaces are confined to the sub-micrometre scale, making geometry-programmed magnetic landscapes a promising platform for future studies of transport and optical phenomena at engineered magnetic interfaces. \\

\noindent\textbf{Conclusions and Outlook}

We have introduced geometronics, a strategy in which geometry serves as an independent degree of freedom for programming the electronic and magnetic landscape of intrinsically anisotropic van der Waals materials. Unlike existing approaches based on chemical composition, electrostatic gating, strain engineering, or heterostructure assembly, geometronics exploits substrate topography to locally control the orientation of the crystal with respect to a spatially uniform external field, thereby creating spatial variations in its electronic or magnetic structure.

As a proof of concept, we demonstrated this principle using a bilayer of antiferromagnetic CrSBr conformally transferred onto an inverted pyramidal nanoindentation. The local variation of crystal orientation converts a spatially uniform magnetic field into spatially varying antiferromagnetic and ferromagnetic phases that coexist within a single continuous crystal. This coexistence creates a reconfigurable geometry-defined excitonic potential well with a depth reaching 10--12~meV that follows the engineered substrate topography. In principle, even larger potential depths should be achievable through further optimisation of the substrate geometry and the local crystal orientation. Together, these results establish geometry-controlled electronic structure engineering as a viable strategy for creating programmable excitonic landscapes in anisotropic layered materials.

Beyond providing a new route for engineering excitonic landscapes in CrSBr, our approach enables the creation of interfaces between distinct magnetic phases. Such interfaces can provide an attractive platform for investigating the coupled transport and dynamics of charge carriers, excitons and magnons across antiferromagnetic--ferromagnetic boundaries, where their mutual interactions are expected to produce rich emergent phenomena. More generally, geometronics offers a route to heterostructure-like electronic landscapes on-demand within a single, compositionally and structurally homogeneous crystal, without interfaces, junctions, or fabrication steps beyond substrate design.

Crucially, the underlying principle relies only on the coexistence of mechanical flexibility and anisotropic material properties and is therefore broadly applicable to many layered materials. While demonstrated here through the strong coupling between magnetic order and electronic structure in CrSBr, the same concept can be extended to any system exhibiting anisotropic magnetic, electronic or excitonic responses. Potential implementations include geometry-programmed energy landscapes based on anisotropic Landé $g$-factors \cite{robert2020measurement, dyksik2021brightening}, Stark shifts \cite{scharf2016excitonic, pedersen2016exciton}, ferroelectric polarisation \cite{wang2021interfacial,deb2024excitonic}, or other orientation-dependent interactions. We therefore anticipate that geometronics will establish substrate topography as a fundamental design principle for programming on-demand tunable and local electronic and magnetic functionalities in anisotropic van der Waals materials.

\section*{Methods}

\subsection{Sample synthesis}

CrSBr single crystals were synthesized following the procedure described in Ref.~\cite{smiertka2026distinct}.

\subsection{Sample preparation}

Nanoindentation pits were fabricated in the Si/SiO$_2$ substrate with the oxide thickness of 285 nm using a Vickers nanoindentation tester (NHT, CSM Instruments, NHT2) equipped with a three-sided pyramidal Berkovich tip, producing an array of inverted pyramidal pits with a lateral size of 4.8~$\mu$m and a depth of 320~nm. 

CrSBr bilayers were obtained by mechanical exfoliation from bulk crystals. The crystals were first thinned using blue Nitto tape (Nitto Denko), after which the exfoliated material was transferred onto a polydimethylsiloxane (PDMS) stamp (Gel-Pak) mounted on a glass slide. Bilayer flakes were identified on the PDMS using an optical microscope (HQ
Graphene) with a camera (Thorlabs, Zelux).  Selected CrSBr bilayers were then deterministically transferred onto substrates containing nanoindentation pits using an all-dry viscoelastic stamping technique~\cite{castellanos2014deterministic}. The glass slide carrying the PDMS stamp and the selected flake was aligned with the target nanostructure and brought into contact with the substrate using a Z-axis micromanipulation stage (Thorlabs Z825B). The stamp was subsequently retracted slowly, leaving the flake on the substrate.

\subsection{Optical spectroscopy}

Low-temperature micro-photoluminescence ($\mu$PL) measurements were performed in the Faraday configuration, with the magnetic field applied perpendicular to the sample plane. The experiments were carried out using a free-beam optical setup integrated with a superconducting magnet capable of generating magnetic fields of up to \(16~\mathrm{T}\).

The sample was mounted on an \(x\)-\(y\)-\(z\) piezoelectric stage and maintained at a temperature of \(10~\mathrm{K}\) in a helium-gas atmosphere. The spatial resolution of the measurements was approximately \(1~\mu\mathrm{m}\).

For PL measurements, the sample was excited using a continuous-wave (CW) laser diode with a wavelength of \(660~\mathrm{nm}\) (\(1.88~\mathrm{eV}\)), coupled through an optical fibre. The excitation beam was focused onto the sample using a \(100\times\) microscope objective (\(\mathrm{NA}=0.82\)), resulting in a spot size of approximately \(1~\mu\mathrm{m}\) in diameter. The emitted signal was collected through the same objective, dispersed by a spectrometer with a focal length of \(0.50~\mathrm{m}\), and detected using a CCD camera.

Spatial line mapping was automated using a custom control script. Starting from a designated initial position, line scans were acquired along the \(y\)-axis over a total distance of 50 $\mu$m in 100 discrete steps. At each step, the piezoelectric stage translated by a fixed increment prior to initiating spectrum collection. Restricting movement to a single axis of the stage during these measurements minimised mechanical drift, ensuring higher spatial precision and reproducibility than two-axis spatial mapping. 

\subsection{Topography characterization}
Atomic force microscopy measurements were performed using a Nanosurf FlexAFM system controlled by a Nanosurf C3000 controller. A BudgetSensors Tap150Al-G cantilever was operated in phase-contrast tapping mode. The measurements were carried out over \(20 \times 20~\mu\mathrm{m}^2\) scan areas with a resolution of \(2048\) pixels.

\section*{Data availability}
All data supporting the findings of this study are available on Zenodo at https://doi.org/10.5281/zenodo.22226777

\section*{Acknowledgements}
PP and MS acknowledge the National Science Centre
Poland within the Opus Program (Grant number 2025/57/B/ST3/03197). MB and EC acknowledge the National Science Centre
Poland within the program Sonata Bis (2020/38/E/ST3/00194).
The work was created as part of a project co-financed by the Polish Ministry of Science and Higher Education under Contract No. 2025/WK/01.
Z.S. was supported by ERC-CZ program (project LL2101) from Ministry of Education Youth and Sports (MEYS) and by the Advanced Multiscale Materials for Key Enabling Technologies project, supported by the Ministry of Education, Youth, and Sports of the Czech Republic. Project No. CZ.02.01.01/00/22\_008/0004558 , Co-funded by the European Union.

\section*{Author contributions}
M.Ś. performed the optical and magneto-optical measurements, analysed the data, prepared the initial draft of the manuscript, and contributed to the preparation of figures for the main text and Supplementary Information. E.C. prepared the sample, conducted the topographic characterisation and additional optical and magneto-optical measurements, prepared figures for both the main text and Supplementary Information, and contributed to the data analysis and drafting of the Supplementary Information. O.J. performed the optical measurements and contributed to the magneto-optical measurements and drafting of the Supplementary Information. B.H. assisted with the magneto-optical measurements. G. and G.K. supported the magnetic field measurements. M.H. fabricated the $\mathrm{Si/SiO_2}$ substrates featuring inverted pyramidal nanoindentations. K.M. synthesised the CrSBr crystals under the supervision of Z.S. M.R.M. and A.B. supervised the magnetic field measurements. M.B. and P.P. conceived the study and developed its overall concept. They supervised the magnetic field measurements, contributed to the analysis and interpretation of the data, and participated in writing, reviewing, and finalising the manuscript.

\section*{Competing interests}
The authors declare no competing interests.

\bibliography{Bibliography_Experiment}

\end{document}